\documentclass[11pt,twoside,A4]{article} 
\usepackage{times,fancyhdr}
\usepackage[dvips]{graphicx}
\usepackage{latexsym} 
\usepackage[affil-it]{authblk}
\usepackage{amsmath}
\usepackage{amssymb}
\usepackage{setspace}

\usepackage[top=4cm, bottom=4cm, left=4cm, right=4cm]{geometry}

\def\beq{\begin{equation}}
\def\eeq{\end{equation}}
\def\beqn{\begin{eqnarray}}
\def\eeqn{\end{eqnarray}}

\begin{document}
 
\title{Analog Systems for Gravity Duals}
\author{Sabine Hossenfelder\thanks{hossi@nordita.org}} 
\affil{\small Nordita\\
KTH Royal Institute of Technology and Stockholm University\\
Roslagstullsbacken 23, SE-106 91 Stockholm, Sweden}
\date{}
\maketitle
\begin{abstract}
We show that analog gravity systems exist for charged, planar black holes in asymptotic Anti-de Sitter space. These black
holes have been employed to describe, via the gauge-gravity duality, strongly coupled condensed matter systems on
the boundary of AdS-space. The analog gravity system is a different condensed matter system that, in a suitable limit, 
describes the same bulk physics as the theory on the AdS boundary. This combination of the gauge-gravity duality and 
analog gravity therefore suggests a duality between different condensed matter systems. 
\end{abstract}

\section{Introduction}

Analog gravity \cite{Unruh:1980cg,Barcelo:2005fc} is the use of condensed-matter systems to model (quantum) field theory in curved space.
It is based on the observation that excitations over the background in certain condensed matter systems
fulfill equations of motion that are analytically identical to that of perturbations traveling in a 
space-time background with a metric different from Minkowski space. The perturbations can
be classical or quantum. 
In the quantum case, analog
gravity makes it possible to experimentally test quantum field theory in curved backgrounds, which is
presently not possible in the actual space-time background due to the weakness of the
gravitational interaction.

In the condensed-matter
system of analog gravity, the effective metric is created by the flow and density of the background. This restricts
the class of space-times that can be simulated because it must be possible to bring the metric into
a certain form. This form differs depending on whether the condensed matter system is
relativistic or non-relativistic, classical or quantum. Spacetimes that can be simulated with gravitational analogs
include the Schwarzschild black hole \cite{Visser:1997ux,Garay:1999sk,Barcelo:2000tg} and expanding de-Sitter space that mimics
the inflationary epoch of the early universe \cite{Volovik:2000ua,Weinfurtner:2004mu,Jain:2007gg,Lin:2012ft,Bilic:2013qpa}. 
Analog gravity is presently a very lively research area in which both the theory and the experiments are being
rapidly developed \cite{Weinfurtner:2010nu,ex1}. In analog gravity, the equations of motion of the background
itself are those of the analog gravity system and will not in general reproduce Einstein's field
equations for this would require full background independence. 



The gauge-gravity correspondence relates a strongly coupled conformal or near-conformal
quantum field theory ({\sc CFT}) to gravity in Anti-de Sitter (AdS) space \cite{Maldacena:1997re,Witten:1998qj,Gubser:2002tv}. The quantum field theory is a theory
on the boundary of the AdS space and thus in a space with one spatial dimension less than the AdS space. This
AdS/{\sc CFT} correspondence has received much attention as a possible way to tackle strongly
coupled systems that with other methods are technically very difficult to treat, for example the quark gluon
plasma and strange metals near quantum criticality. The latter case that we are
interested in here is also sometimes
referred to as AdS/{\sc CMT} duality, where {\sc CMT} stands for Condensed Matter Theory. 

The best studied case of AdS/CMT is the phase transition to superconductivity in
high-temperature superconductors, or their dual gravitational description respectively.
The gravitational dual that serves as model for the superconductor is taken to be a charged black hole in AdS space. 
In this background propagates a massive scalar field coupled to a U(1) gauge
field. The temperature of the black hole corresponds to the temperature of the
material in the theory of the boundary. The additional fields are necessary to induce 
an instability at a transition temperature at which the fields condense around
the black hole. By studying perturbations traveling in these backgrounds one
can calculate properties of the condensed-matter medium, such as its
electric conductivity or optical resistance \cite{Hartnoll:2008kx,Hartnoll:2009sz,Horowitz:2010gk,Musso:2014efa}. So
far the results are qualitative, but the method is being rapidly developed and
has a large potential to significantly advance theoretical physics.

In principle the equations of the perturbation in the background of the AdS black
hole are coupled to the equations that determine the background metric. In
practice however the calculations are often performed in the `probe approximation'
in which the backreaction of the fields on the metric is negligible. This 
approximation has been found to capture much of the interesting physics \cite{Horowitz:2010gk}, and this
means that one often actually deals with quantum field theory in a curved background.

This opens the intriguing possibility that it might be possible to create an
analog gravity system of the dual theory, thereby relating the 
strongly coupled condensed-matter system on the (flat) boundary of AdS space
to a weakly coupled condensed-matter system also in flat space, but generating
an effective metric that reproduces the asymptotical AdS geometry. This relation should exist if 
it is possible to model AdS planar black holes with analog gravity. 

In this
paper, we will take the first, and most essential step towards studying the
possibility that there exists a new duality between condensed matter systems,
that is we will show that the metrics in the gravitational dual of holographic
superconductors are indeed effective analog metrics of certain types of
fluids.

\section{Motivation}

Some dualities have been known
for a long time \cite{Polchinski:2014mva}, but it has only been in the last two decades that
we have begun to understand the full relevance of dualities. Not only are dualities that relate
a strongly with a weakly coupled theory useful computational tools. They also make us
reconsider what really is fundamental about a theory, as we see now that we have a choice to
select the fields and their symmetries from either side of the duality. Whichever side we take to be fundamental, the other
side will appear to have emergent features. String theory in particular is known to have
a web of dualities that is believed to be completed by a so-far little understood theory called
M-theory. Exploring dualities is thus
both of practical value as well as pushing ahead on questions pertaining unification and
 the fundamental nature of physical theories.

Besides better understanding the relations between physical laws in general, 
finding out whether there exist so-far unknown relations between two condensed matter systems 
by combining the AdS/{\sc CMT} correspondence with analog gravity is interesting for
the following three reasons. 

First, such a relation between
strongly and weakly coupled quantum field theories, even if valid only in some limit,
could be helpful to understand the theory of
strange metals by identifying observables of the one system with observables of the
other system. If such a relation exists, it means that some behaviors of these systems
can be mapped to each other, and so known insights about one class of systems can reveal
new insights about the other class of systems.

Second, while the gravitational dual of the strongly coupled
field theory simplifies the calculation by avoiding the strong coupling regime, one 
instead has to deal with several coupled non-linear differential equations that
are not analytically solvable unless in cases with many symmetries. Realistic
high-temperature superconductors however have quite complicated multi-layered
lattice symmetries whose gravitational dual is difficult to study even numerically,
though some progress has been made in recent years for simple lattice symmetries \cite{Horowitz:2012ky,Horowitz:2013jaa}. 
An analog dual of the superconductor would open the possibility to 
directly simulate the system by means of another system and measuring rather
than compute the quantities, thus shortcutting the numerics. For this, one
would have to show that not only does the analog gravity system reproduce
the field theory in the asymptotic AdS space, but that also the backreaction
is properly taken into account. In this paper, we will only look at the the
fixed background case, which is a necessary condition that must be fulfilled
if the full dynamics also can be mapped. 

Third, the identification between two condensed matter systems that is
based on combining the AdS/{\sc CMT} correspondence with analog gravity can
be used as a way to experimentally test the correspondence. The best understood case of the gauge-gravity duality is the AdS$_5$/CFT$_4$ duality \cite{Maldacena:1997re,Witten:1998qj,Gubser:2002tv},
a correspondence between Type IIB Superstring theory on AdS$_5 \times$S$^5$ and the maximally supersymmetric
SU(N) Yang-Mills theory in four dimensions. There are also some other
examples of AdS/{\sc CFT} dualities, which are lower-dimensional and less supersymmetric. But for most of 
the theories used in the description of holographic superconductors it is not presently known whether the gauge 
dual exists, though recent
progress has been made towards answering this question \cite{ElShowk:2011ag,Fitzpatrick:2012cg}. 
If it can be demonstrated that the relation between the two condensed matter systems that
we will investigate in this paper is a correct description of nature, then this would imply that the
AdS/{\sc CMT} correspondence used to obtain this relation is also correct. 

Recently it has been demonstrated that analog gravitational systems for
general Fried\-mann-Robertson-Walker ({\sc FRW}) spaces, including AdS, exist \cite{Bilic:2013qpa} though
these have not yet been experimentally realized. AdS space has many
symmetries and this gives one hope that it may not be too difficult to
model, and probably the general {\sc FRW} case discussed in \cite{Bilic:2013qpa} is
not necessary. On the other hand we need black holes in AdS space,
which goes beyond the gravitational analogs studied in the literature
so far.

\begin{figure}[ht]
\includegraphics[width=9.5cm]{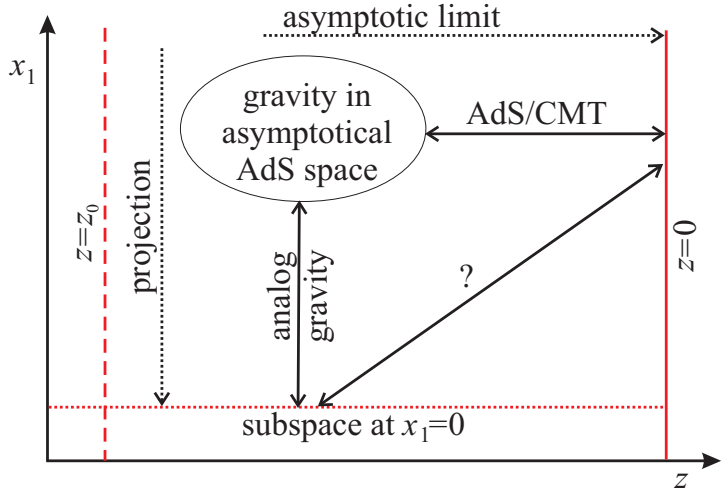} 

\caption{Sketch of relations between different expressions of the same theory. Shown
is the asymptotic coordinate, $z$, of AdS-space and one of the perpendicular coordinates $x_1$.
We will assume that we have a translational symmetry in $x_1$-direction and will here only
consider the static case, thus the time-coordinate is not shown. The boundary of AdS-space is at $z=0$, where we have a strongly
coupled condensed matter system.
The AdS/{\sc CMT} correspondence relates this boundary theory to a gravitational
theory in the bulk which contains fields propagating in the curved background. The
same gravitational system can alternatively be described as an analog gravity system
in which the metric in AdS space is only emergent for weakly coupled 
perturbations propagating in some fluid's background. Projected onto a subspace, one
now has two condensed matter systems in flat space-time of the same dimension that describe the same physics,
if in entirely different ways. What is the relation between these two theories? \label{fig1}}
\end{figure}

It has previously been proposed \cite{Bilic:2014dda} to study the gravity dual of a condensed
matter system on the boundary which is also an analog gravity system. This is not what we
will be doing here. We will look for an analog gravity system {\it in the bulk}, not on the
boundary. Since the space-time in the bulk then has one additional dimension
that cannot be simulated in the lab if the boundary is already 3+1 dimensional,
we will in the end be interested in a spatial slice of AdS space which has the
same dimension as the AdS-boundary, but the projection is onto a coordinate
perpendicular to the asymptotic coordinate (see figure \ref{fig1}). 

This then means that we have two condensed matter systems in flat space,
one on the boundary of AdS, and one on a spatial slice perpendicular to the boundary. The former
condensed matter theory is strongly
coupled, the latter weakly coupled. The condensed matter system on the AdS
boundary is related to the gravity in the bulk through the AdS/CMT correspondence,
and the gravity in the bulk is then identified with a different condensed matter system
by using the gravitational analog. At least in the probe limit and in the limit of small
perturbations these relations are
exact -- provided we can show that there is a gravitational analog for interesting
geometries that are asymptotically AdS to begin with. These two identifications, if possible, then would mean that
the strongly coupled condensed matter theory on the AdS boundary must be
related to the weakly coupled condensed matter system that reproduces the
AdS subspace. If true, this is a surprising relation worth further investigation.

In the following the signature of the metric is $(-1,1,1...)$. Small Greek indices run over all space-time coordinates, small Latin
indices run over spatial coordinates only. The speed of light and $\hbar$ are equal to one. $c$ is
the speed of sound, {\it not} the speed of light. 

\section{Analog Gravity for a classical fluid}

We start with briefly summarizing how one derives the effective metric of the analog gravity systems. The
reader already familiar with analog gravity may skip to section \ref{4}.

\subsection{Non-relativistic case}

We here follow the treatment of \cite{Barcelo:2005fc}. The dynamics of a fluid with vanishing
viscosity is determined by the continuity equation and the Euler equation:
\beqn
\partial_t \rho + {\vec \nabla} \cdot (\rho {\vec v} ) &=& 0 \label{continuity} ~, \\
\rho \left[ \partial_t {\vec v} + ( {\vec v} \cdot {\vec \nabla})  {\vec v} \right] &=& \vec{F} \label{euler}~,
\eeqn
where $\rho$ is the density, ${\vec v}$ is the 3-vector describing the fluid's velocity field, and ${\vec F} = - {\vec \nabla} p$ is a force field
generated by the pressure $p$.
We will in the following assume that the fluid is vorticity free, ie locally irrotational. In this case, the velocity field can
be expressed as the gradient of a scalar field ${\vec v} = - {\vec \nabla} \phi$. We will also assume that the fluid is
barotropic, so that the density $\rho$ is a function of $p$ only. Then one can express the specific enthalpy $h(p)$ as
\beqn
h(p) = \int_0^p \frac{dp'}{\rho(p')} ~,
\eeqn
which, taking the derivative, implies
\beqn
{\vec \nabla} h = \frac{1}{\rho} {\vec \nabla} \rho ~.
\eeqn

In the vorticity-free case Euler's equation (\ref{euler}) can then be rewritten as
\beqn
\partial_t {\vec v} = - \frac{1}{\rho} {\vec \nabla} p - \frac{1}{2} {\vec \nabla} \left(  v^2 \right)~, \label{euler2}
\eeqn
where $v$ is the absolute value of ${\vec v}$. Using the definitions for $\phi$ and $h$, equation (\ref{euler2})
can be integrated once and
takes on the form
\beqn
 \partial_t \phi =  h + \frac{1}{2} \left( \vec \nabla \phi \right)^2~. \label{euler3}
\eeqn

The whole task of analog gravity is now to show that with a suitable decomposition of these equations,
a perturbation over the fluid fulfills a wave-equation in an effective background metric that is 
determined by the fluid. For this we make an expansion around the background described by $\rho_0, p_0, \phi_0$
\beqn
\rho &=& \rho_0 + \varepsilon \rho_1 + {\cal O}(\varepsilon^2) ~, \\
p &=& p_0 + \varepsilon p_1 + {\cal O}(\varepsilon^2) ~,\\
\phi &=& \phi_0 + \varepsilon \phi_1 + {\cal O}(\varepsilon^2) ~.
\eeqn
(To avoid confusion, we will make sure in the following to use the numbers $0$ and $1$ only to label the fields.
If the indices refer to coordinates we will use the name of the coordinates instead.)
Taking apart the continuity equation (\ref{continuity}) into the zeroth and first order terms yields
\beqn
\partial_t \rho_0 + {\vec \nabla} \cdot \left( \rho_0 {\vec v}_0 \right) &=& 0 ~, \label{continuityback} \\
\partial_t \rho_1 + {\vec \nabla} \cdot \left( \rho_1 {\vec v}_0 + \rho_0 {\vec v}_1 \right) &=& 0 ~, \label{continuitypert}
\eeqn
where ${\vec v}_0 = - {\vec \nabla} \phi_0$ and ${\vec v}_1 = - {\vec \nabla} \phi_1$, respectively. 

The  specific enthalpy can be expanded as
\beqn
h(p) = h(p+ \varepsilon p_1 + {\cal O}(\varepsilon^2) ) = h_0 + \varepsilon \frac{p_1}{\rho_0} + {\cal O}(\varepsilon^2) ~. \label{exh}
\eeqn
With (\ref{exh}) the integrated Euler equation (\ref{euler3}) can be taken apart into a zeroth and first order term to give
\beqn
\partial_t \phi_0 &=& h_0 + \frac{1}{2} \left( {\vec \nabla} \phi_0 \right)^2 ~, \\
\partial_t \phi_1 &=& \frac{p_1}{\rho_0} - {\vec v}_0 \cdot {\vec \nabla} \phi_1  ~. \label{phi1}
\eeqn

We now have two equations each for the background and for the perturbation. Next we direct our attention to the scalar
field $\phi_1$. The equation determining its dynamics (\ref{phi1}) can be rewritten as
\beqn
p_1 = \rho_0 \left( \partial_t \phi_1 + {\vec v}_0 \cdot {\vec \nabla} \phi_1 \right) ~.
\eeqn
In the limit of small perturbations we have $\partial \rho/\partial p \approx \rho_1/p_1$ and thus
\beqn
\rho_1 = \frac{\partial \rho}{\partial p}\rho_0 \left( \partial_t \phi_1 + {\vec v}_0 \cdot {\vec \nabla} \phi_1 \right) ~. \label{rho1}
\eeqn
We can then use (\ref{rho1}) to eliminate $\rho_1$ from equation (\ref{continuitypert}) to get
\beqn
\partial_t \left( \frac{\partial \rho}{\partial p} \rho_0 \left( \partial_t \phi_1 + {\vec v}_1 \cdot {\vec \nabla} \phi_1 \right) \right) 
- {\vec \nabla} \left (\rho_0 {\vec \nabla} \phi_1 - \frac{\partial \rho}{\partial p} \rho_0 {\vec v}_0 
\left( \partial_t \phi_1 + {\vec v}_1 \cdot {\vec \nabla} \phi_1 \right) \right) = 0 ~. \label{phiwave}
\eeqn
This expression looks somewhat messy, but the important thing to note is that it is a linear second order equation for
$\phi_1$ with coefficients that depend only on the background fluid $(\rho_0, \vec v_0)$. With the definition $c^{-2} = \partial \rho_0/\partial p_0$ and
\beqn
g^{\mu \nu} (t, {\vec x}) \propto \frac{1}{c \rho_0}
\left( \begin{array}{cc}
-1 & -v_0^j \\
-v_0^i & (c^2 \delta^{ij} - v_0^i v_0^j ) \end{array} \right)~, \label{gup}
\eeqn
the equation (\ref{phiwave}) be equivalently written as
\beqn
\partial_\mu \left( \sqrt{- g} g^{\mu \nu} \partial_\nu \phi \right) =0~, \label{weq}
\eeqn
which is the wave-equation of a scalar perturbation in a curved background with metric (\ref{gup}). The
physical interpretation of $c$ is the local speed of sound in the background. This analog metric might have a constant prefactor that
does not affect the equation of motion and therefore cannot be obtained by reading off the metric from the wave-equation. 
We will identify this prefactor later from the Lagrangian approach.

It should be pointed out that the metric (\ref{gup}) is just one example for an analog metric. It is
in the literature often referred to as `acoustic metric' to indicate that it deals specifically with sound
propagation. 

The inverse of the inverse metric (\ref{gup}) is the metric
\beqn
g_{\mu \nu} (t, {\vec x}) \propto \frac{\rho_0}{c}
\left( \begin{array}{cc}
-(c^2-v_0^2) & -v_0^j \\
-v_0^i &  \delta_{ij}  \end{array} \right)~, \label{gdown}
\eeqn
where $v_0^2 = \sum_i v_0^i v_0^i$. The line-element then reads
\beqn
{\rm d}s^2 \propto \frac{\rho_0}{c} \left( - c^2 {\rm d}t^2 +\delta_{ij} ({\rm d}x^i - v_0^i {\rm d} t)({\rm d}x^j - v_0^j {\rm d} t) \right) ~.
\eeqn

Let us sit back for a moment and look at this derivation backwards. Suppose we start with a metric rather
than with the fluid, as we will be doing in the later sections. We then use the expression (\ref{gup}) to identify the degrees of
freedom of the fluid from the metric. Then we have to check that these quantities actually fulfil the equations
of motion of the fluid. The important constraint comes from the continuity equation. The Euler-equation can
always be made to be fulfilled by adding a suitable external force. 

Note that the derivation of the wave-equation from the fluid-dynamics is independent of the number
of dimensions. All that changes with adding spatial dimensions is the number of indices that are being summed over. However, the
reformulation in terms of propagation in a non-flat background depends on the number of dimensions
because the wave-equation (\ref{weq}) contains the metric determinant to reproduce the covariant
derivative. The power of the metric's prefactor in the determinant depends on the number of (non-degenerate)
dimensions in the metric and this power has to be correct so as to reproduce (\ref{phiwave}) when multiplied
with the metric itself. For a system with $n$ spatial dimensions the scaling of the effective acoustic metric 
and its inverse is thus
\beqn
g^{\mu \nu} (t, {\vec x}) &\propto& \left( \frac{\rho_0}{c} \right)^{-\frac{2}{n-1}}
\left( \begin{array}{cc}
-1/c^2 & - v_0^j/c^2 \\
-v_0^i/c^2 & \delta^{ij} - v_0^i v_0^j/c^2  \end{array} \right)~, \label{gupd} \\
g_{\mu \nu} (t, {\vec x}) &\propto& \left( \frac{\rho_0}{c} \right)^{\frac{2}{n-1}}
\left( \begin{array}{cc}
-(c^2-v_0^2) & - v_0^j \\
-v_0^i & \delta_{ij}   \end{array} \right) \label{gdownd} ~.
\eeqn

Note that this scaling only applies for systems confined to $n+1$ dimensional space, not to
higher dimensional systems that have a symmetry which effectively reduces the degrees of
freedom to $n+1$ dimensions. That is, if we have a fluid with excitations confined to a $k$-dimensional surface, then we 
use $n=k$. But if we have an $n$-dimensional system that has a planar symmetry in $k$ of the $n$ dimensions and
we want to model the planar subspace, we would still use $n$ in the scaling rather than $k-n$.

\subsection{Relativistic case}

The analog metric of a relativistic fluid was derived in \cite{Bilic:1999sq,Visser:2010xv}. We here follow the derivation of
\cite{Barcelo:2005fc,Bilic:2013qpa} using a Lagrangian approach. We reproduce part of this derivation because
we will use it later to add the U(1) gauge field. 

Normally one works with hydrodynamics on the level of the
equations of motion. One can then construct a Lagrangian that gives rise to these equations of motion, but
if one has the equations already it is somewhat pointless to make this effort, so it is rarely done. However, in our
case it will be beneficial to see how the effective background metric comes about. The general Lagrangian of a fluid
$\cal L$ depends on a kinetic energy term $\chi$ of the form 
\beqn
\chi = \eta^{\mu\nu} \left( \partial_\nu \theta \right) \left( \partial_\mu \theta \right)
\eeqn
and some additional potential $V(\theta,t,\vec x)$ that we take to include a possible mass term. This potential will in general be non-trivial and not just be a mass-term because we are dealing with a
condensed matter system that we typically want to, well, condense, around some minimum of the potential. We will
select a specific potential later in section \ref{mass} but keep it general for now. The full Lagrangian then
has the form
\beqn
{\cal L} = {\cal L}(\chi(\partial \theta) - V(\theta, t, \vec x)) ~.
\eeqn

The energy momentum-tensor for this Lagrangian can be obtained by variation
with respect to the metric and is
\beqn
T_{\mu\nu}= - \left( 2 \frac{\partial {\cal L}}{\partial \chi} (\partial_\nu \theta)( \partial_\mu \theta) - {\cal L} \eta_{\mu \nu} \right) ~.
\eeqn
One can then identify pressure $p_0$, density $\rho_0$ and the fluid-velocity $u_\nu$ by comparing the stress-energy
derived from the Lagrangian to the familiar stress-energy tensor of a fluid
\beqn
T_{\mu \nu} = (p_0+\rho_0)u_\mu u_\nu + p_0 \eta_{\mu \nu} ~.
\eeqn
So one finds 
\beqn
u_\nu = \frac{\partial_\nu \theta}{\sqrt{\chi}}~,~p_0 = {\cal L}~,~\rho_0 = 2 \chi \frac{\partial {\cal L}}{\partial \chi} - {\cal L}~, \label{ident}
\eeqn
where the four-velocity is normalized to one
\beqn
\eta^{\mu\nu}u_\mu u_\nu = - 1~.
\eeqn
The field equations of the relativistic fluid are the conservation of the stress energy
\beqn
\partial_\nu T^{\mu \nu} = 0~.
\eeqn
In the non-relativistic limit one reproduces the continuity-equation from the equation with $\mu =0$ and the Euler-equation
from the other three. Again one can add a force field to the right side. The relevant thing to notice here
is that the kinetic part of the Lagrangian of the fluid will not generally be linear in $\chi$
but, depending on the equation of state, ${\cal L} = p_0$ will be some function of $\chi$, so the
second derivative of ${\cal L}$ with respect to $\chi$ will not normally just vanish.

We now make a Taylor-series expansion of the action around a background field $\theta_0$ that is assumed to fulfil
the Euler-Lagrange equations belonging to the unperturbed ${\cal L}$
\beqn
\theta(t,{\vec x}) = \theta_0(t,{\vec x}) + \varepsilon \theta_1(t,{\vec x}) ~. \label{pertan}
\eeqn
The expansion of the Lagrangian is then
\beqn
&&{\cal L}(\theta, \partial_\nu \theta) = {\cal L}(\theta_0, \partial_\nu \theta_0) + \varepsilon \left[ \frac{\partial {\cal L}}{\partial (\partial_\mu \theta)} \partial_\mu \theta_1 + \frac{\partial {\cal L}}{\partial \theta} \theta_1\right] \nonumber \\ 
 &&+\frac{\varepsilon^2}{2} \left[ \frac{\partial^2 {\cal L}}{\partial (\partial_\nu \theta) \partial (\partial_\mu \theta)} (\partial_\nu \theta_1)( \partial_\mu \theta_1) + 2 \frac{\partial^2 {\cal L}}{\partial(\partial_\nu \theta)\partial \theta} (\partial_\nu \theta_1) \theta_1 + \frac{\partial^2 {\cal L}}{\partial \theta \partial \theta} \theta_1 \theta_1\right] ~, \label{beginderiv}
\eeqn
plus terms of order $\varepsilon^3$ and higher. The term of order $\varepsilon$, when plugged into the action and after a partial integration, gives the equation of motion for the background field when assumed to vanish separately, since this is how one normally derives the equation of motion. If the
background field is assumed to fulfil the equations of motion in the non-perturbed case, then this term just drops out. The remaining terms can,
with another partial integration, be
collected to give
\beqn
 S[\theta] &=& \int {\rm d}^{d+1} x {\cal L} = S[\theta_0] + \nonumber \\ 
 \frac{\varepsilon^2}{2} \int {\rm d}^{d+1} x && \hspace*{-1cm}
\left[ \frac{\partial^2 {\cal L}}{\partial (\partial_\nu \theta) \partial (\partial_\mu \theta)} (\partial_\nu \theta_1)( \partial_\mu \theta_1) 
+  \left( \frac{\partial^2 {\cal L}}{\partial \theta \partial \theta} - \partial_\nu \left( \frac{\partial^2 {\cal L}}{\partial(\partial_\nu \theta) \partial \theta}\right) \right) \theta_1 \theta_1 \right].
\eeqn
One sees that the contributions from the perturbations take the form of the action of a scalar field propagating in an effective metric $g_{\mu\nu}$ defined by
\beqn
\sqrt{-g} g^{\mu \nu} = - \frac{\partial^2 {\cal L}}{\partial (\partial_\nu \theta) \partial (\partial_\mu \theta)} ~, \label{geff}
\eeqn
and with a mass term
\beqn
\sqrt{-g} m_{\rm eff}^2 = -  \frac{\partial^2 {\cal L}}{\partial \theta \partial \theta} + \partial_\nu \left( \frac{\partial^2 {\cal L}}{\partial(\partial_\nu \theta) \partial \theta}\right)  ~. \label{meff}
\eeqn
This effective mass 
will not in general be a constant.  

We can now rewrite the partial derivatives using
\beqn
\frac{\partial}{\partial (\partial_\mu \theta)} = \frac{\partial \chi}{\partial (\partial_\mu \theta)} \frac{\partial}{\partial \chi} = 2 \eta^{\mu \nu} \partial_\nu \theta \frac{\partial}{\partial \chi} ~,
\eeqn
then we obtain
\beqn
\sqrt{-g} g^{\mu \nu} = - 2 \left( \eta^{\mu\nu} \frac{\partial {\cal L}}{\partial \chi} - 2 \chi u^\mu u^\nu \frac{\partial^2 {\cal L}}{\partial \chi^2} \right) ~.
\eeqn
Keep in mind that the fluid lives in flat space and the curved background is only emergent, so the indices on $u_\nu$ are raised and lowered with $\eta^{\mu\nu}$ and
$\eta_{\mu\nu}$, respectively.

To get the effective metric, one now has to calculate the determinant of the above expression. One does this most easily by considering the
case in which the fluid is in the rest frame $u_\nu =(1,\vec 0)$. Then one finds in $n+1$ dimensions
\beqn
(-1)^{\frac{n+1}{2}}g^{\frac{n-1}{2}} = 2^{n+1} \left( \frac{\partial {\cal L}/\partial \chi + 2 \chi \partial^2 {\cal L} /\partial \chi^2}{\partial {\cal L}/\partial \chi} \right) \left( - \frac{\partial {\cal L}}{\partial \chi} \right)^{n+1} ~.
\eeqn
Now we get from our earlier identification (\ref{ident}) that
\beqn
\frac{1}{c^2} = \frac{\partial \rho_0}{\partial {\cal L}} = \frac{\partial \chi}{\partial {\cal L} }\frac{\partial \rho_0}{\partial {\cal \chi}} = \frac{2 \chi \partial^2 {\cal L}/\partial \chi^2 + \partial {\cal L}/\partial \chi}{\partial{\cal L}/\partial \chi} ~, 2 \frac{\partial {\cal L}}{\partial \chi} = \frac{\rho_0+p_0}{\chi}~, \label{ident2}
\eeqn
so that
\beqn
\sqrt{-g} = c^{-\frac{2}{n-1}} \left( - \frac{\rho_0+p_0}{\chi} \right)^{\frac{n+1}{n-1}}~.
\eeqn
With this we then get
\beqn
g^{\mu \nu} = c^{\frac{2}{n-1}}  \left(  \frac{\rho_0+p_0}{\chi} \right)^{-\frac{2}{n-1}}  \left( \eta^{\mu\nu} + \left(1- \frac{1}{c^2} \right) u^\mu u^\nu \right) ~,
\eeqn
and
\beqn
g_{\mu \nu} = c^{\frac{-2}{n-1}}  \left(  \frac{\rho_0+p_0}{\chi} \right)^{\frac{2}{n-1}}  \left( \eta_{\mu\nu} + \left(1- c^2 \right) u_\mu u_\nu \right) ~, \label{endderiv}
\eeqn
up to a prefactor from powers of $-1$ which we will disregard because it does not matter for the following. We will from now on use the case $n=3$ and
come back to the general case in the discussion. To connect the Lagrangian approach we have considered here with the notation of \cite{Visser:2010xv}, 
it is $\chi = (\rho_0+p_0)^2/n_0^2$, where $n_0$ is the particle-density of the fluid. 

The benefit of the Lagrangian approach is that it is straight-forward to see how we can deal with gauge invariance. We can use a complex
U(1)-gauged scalar and just
make the minimal coupling as usual by replacing $\partial_\nu \to D_\nu = \partial_\nu + {\rm i} e A_\nu$ in the Lagrangian. Since every partial derivative of
$\theta$ then
has a corresponding term $e A \theta$, we will get gauge-covariant
derivatives in all places where we previously had partial derivatives, and the rest remains the same. The conserved current can be derived from the
Lagrangian as $J^\nu = - \partial {\cal L}/\partial A_\nu$. To prevent any possible confusion, note that $\rho_0$ is not the zero-component
of the current because the energy-density isn't the same as the charge density. We will denote the charge-density as $J_t$. 

When dealing with the complex rather than the real-valued scalar, we have to sum over derivatives with respect to
both the field and its conjugate in the final expressions for the effective metric (\ref{geff}) and effective mass (\ref{meff}). The complex
field brings with it another subtlety that we have to take into account which is that the perturbations around the field $\theta$ and its
conjugate $\theta^*$ are not independent, so we should not treat the perturbations as two different fields in the Lagrangian approach. For the complex
scalar, instead of the ansatz (\ref{pertan}), we use the familiar ansatz
\beqn
\theta(t,{\vec x}) = \theta_0(t,{\vec x})e^{{\rm i} \varphi(t,{\vec x})/\bar \varphi } ~,~   \label{pertan2}
\eeqn
where $\theta_0$ is the background field that fulfills the unperturbed equations of motion and $\varphi(\vec x, t)$ is the perturbation whose equation of motion we are interested in. 
The quantity $\bar \varphi$ is a constant normalization factor so that the field $\varphi$ has dimension of energy.
The expressions (\ref{geff}) and (\ref{meff}) then each get a factor $(\theta_0/\bar \varphi)^2$.

\subsection{Non-relativistic limit}

The non-relativistic limit, in which both the speed of the fluid as well as the speed of sound are much smaller than
the speed of light, has been derived in \cite{Visser:2010xv} and was shown to reproduce (\ref{gup}) and (\ref{gdown}), as
expected. 
The physical interpretation of $\chi = (\rho_0+p_0)^2/n_0^2$ is the specific enthalpy, which we had already met in the
non-relativistic limit. In the non-relativistic limit 
$\rho_0+p_0 \to m n_0$, and \cite{Visser:2010xv}
\beqn
\chi \to {\mbox{constant}} := m^2 a^2~,  \label{nrl}
\eeqn
where $m$ is the (effective) mass of the particles of background fluid,
and $a$ is an amplitude (of dimension energy).
This fixes the missing prefactor in the non-relativistic limit to $(ma)^{4/(n-1)}$. The analog metric is then
dimensionless. 

To obtain this limit we assume that the kinetic energy $(\vec \nabla \theta)^2$ is much smaller than the potential energy $V(\theta, t, \vec x)$,
and that the field sits close to the ground state. The equation of motion for the field then just gives, as usual, $\partial V/\partial \theta = 0$, which
will be fulfilled for some value of $\theta_0$ that we will denote $\langle \theta_0 \rangle$. For $V(\theta, \vec x, t) = V(\theta)$, this ground state
will be a position-independent constant, and then
\beqn
\chi \to V(\langle \theta_0 \rangle) = m^2 a^2~.
\eeqn
This necessitates that the potential has a minimum to begin with, and if that minimum is at zero then the non-relativistic
limit is generally ill-defined. This shouldn't surprise us because we need some background condensate around which to define perturbations.
We will come back to this potential in section \ref{mass}, when we discuss the mass terms. 

Since $\bar \varphi$ was an arbitrarily chosen constant,
we now set it to $\bar \varphi = \langle \theta_0 \rangle$, so that the effective metric and effective mass keep the expressions we
derived in (\ref{geff}) and (\ref{meff}). In the remainder of the paper we will use the non-relativistic limit of the acoustic metric.

\section{Analog Gravity for the Schwarzschild Black Hole}

To see the general idea of analog gravity and its uses, let us first look at the Schwarzschild black hole.
The line-element of the
uncharged, non-rotating black hole in $3+1$ dimensions in the
Schwarzschild coordinates is
\beqn
{\rm d} s^2 &=& - \gamma {\rm d}t^2 +  \gamma^{-1} {\rm d}r^2 + r^2 \left( {\rm d} \theta^2 + \sin^2 \theta {\rm d} \phi^2 \right) ~,\\
{\mbox{with}}&& \gamma(r) = 1- \frac{2M G}{r} ~.
\eeqn

The metric in these coordinates cannot represent a non-relativistic acoustic metric because the spatial subspaces are not flat. We can
however make the subspaces flat on the expense of giving up the orthogonality of the time-slicing by using the ansatz
$t=t'+f(r)$. One then readily finds
\beqn
f(r) =  \int {\rm d} r \frac{\sqrt{1- \gamma(r) }}{\gamma(r)}~, \label{genf}
\eeqn
so that the Schwarzschild metric takes the form
\beqn
{\rm d} s^2 = - \gamma {\rm d}t'^2 +  \sqrt{ \frac{2M G}{r}} {\rm d} t' {\rm d}r + {\rm d} r^2 + r^2 \left( {\rm d} \theta^2 + \sin^2 \theta {\rm d} \phi^2 \right) ~. 
\eeqn
These coordinates are known as the Painlev\'e-Gullstrand coordinates \cite{PG1,PG2,PG3}. We now can read
off from the $rr$-component by comparing with (\ref{gdown}) that $c = \rho_0$ and from the $it$-components that  $v_0 = \sqrt{2MG/r}$.
One can then extract from the $tt$-component that $c^2 = 1 $ and so
\beqn
\rho_0 v_0 \sim \sqrt{\frac{2MG}{r}} ~.
\eeqn
This however does not fulfill the continuity equation (\ref{continuityback}) because this would necessitate that
(in static, spherical coordinates) $\rho_0 v_0 \sim 1/r^2$. 
We thus see that while we can express the Schwarzschild-metric in the right form of an analog metric,
there is no fluid that can produce this metric because the continuity equation is not fulfilled.

However, one can find analog gravity systems that produce a metric which is conformally equivalent
to the Schwarzschild-metric in Painlev\'e-Gullstrand coordinates, for example by choosing $v_0 = \sqrt{2MG/r}$, $\rho = r^{-3/2}$ and
$c=$constant. In this case $v_0 \rho_0 \sim r^{-2}$ and the continuity equation is fulfilled. The metric one obtains 
this way is
\beqn
{\rm d}s^2 \propto \frac{1}{r^{3/2}} \left[ - \gamma {\rm d}t'^2 +  \sqrt{ \frac{2M G}{r}} {\rm d} t' {\rm d}r + {\rm d} r^2 + r^2 \left( {\rm d} \theta^2 + \sin^2 \theta {\rm d} \phi^2 \right) \right] ~,
\eeqn
where the constant of proportionality depends on the speed of sound. Perturbations travelling in this analog 
metric will reproduce all the characteristic features of perturbations
in the black hole background, except for the overall scaling which is different due to the conformal factor.

The purpose of this little exercise is to demonstrate that the metric of the Schwarzschild black hole
does not automatically allow a description in terms of the analog metric.

\section{Anti-de Sitter Space}

\label{4}

\subsection{Empty AdS}

Anti-de-Sitter (AdS) space is conformally flat. In $d$+1 dimensions its line element can be written in the form
\beqn
{\rm d}s^2 = \frac{L^2}{z^2} \left( - dt^2 + dz^2 + \sum_{i=1}^{d-1 } {\rm d}x^i {\rm d} x^i  \right) \quad. \label{ads}
\eeqn
Here, $z$ is the asymptotic coordinate that goes to zero at the boundary of AdS space. These
coordinates are commonly used for calculations within the gauge-gravity correspondence because
they make the expansion around the boundary intuitive. $L$ is the length scale of AdS space and related to the
cosmological constant by $\Lambda = -(d-1)(d-2)/L^2$. 

In anticipation of things to come,
we will rescale the time-coordinate by a constant $t \to t \kappa$. We do this because we have set the speed of light
equal to one, but we expect that in the analog system it is instead the speed of sound that plays a similar role\footnote{Keep in mind that the analogy is not generally covariant. By introducing $\kappa$ we generate a family of analog metrics
that will allow us later to identify $\kappa$ from the speed of sound. If the analogy was covariant, we would also have to change the
coordinate system for the real metric and the rescaling would be meaningless.}. This rescaling
gives us the metric
\beqn
{\rm d}s^2 = \frac{L^2}{z^2} \left( - \kappa^2 dt^2 + dz^2 + \sum_{i=1}^{d-1 } {\rm d}x^i {\rm d} x^i  \right) \quad. \label{adsk}
\eeqn

We can then read off the analog gravity metric from Eq (\ref{ads}). Since the metric is diagonal, we just set ${\vec v}_0 = 0$. From
the $g_{ii}$-components, we then get $(\partial {\cal L}/\partial \chi)/c = \rho_0/(m^2 a^2 c) = L^2/z^2$, and from $g_{tt}$ we see that $c^2 =\kappa^2$, so that $\rho_0 = L^2 m^2 a^2 \kappa/z^2$
and the continuity equation is trivially fulfilled. We can then calculate the force-density necessary to create the right pressure
profile by using
\beqn
- c^2 \partial_z \rho_0 = - c^2 \partial_z p_0 \frac{\partial \rho_0}{\partial p_0} = {F_z} ~.
\eeqn
For the empty AdS metric, one finds that the pressure has to fall as $1/z^2$ and the force density 
for empty AdS space in a 3+1 dimensional analog is
\beqn
F_{\rm AdS} := \kappa^3 (L m a)^2 \frac{2}{z^3}~. 
\eeqn
One verifies easily that the mass dimension of the force-density comes correctly to 5.

The reason we find a force term here is that we have not included the full Lagrangian (or its relativistic limit respectively) in the
calculation of the net force. The {\it full} stress-energy must be conserved without the need to add a balancing
force term because it originates in the potential.  We already noted previously that the Euler-equation does not give an additional constraint, and we can see
this here by noting that in the non-relativistic limit the expansion of the Lagrangian for $(\partial \theta_0)^2 \ll V$ is 
\beqn
{\cal L}( \chi  - V(\langle \theta_0 \rangle)) \approx {\cal L}(\chi(v_0 = 0) - V(\langle \theta_0 \rangle)) + \frac{\partial {\cal L}}{\partial \chi}\bigg|_{v_0=0} \chi \theta_0^2 (v_0)^2~. 
\eeqn
This then means that 
\beqn
\partial_z ( \rho v_0^2  - {\cal L} ) = 0~
\eeqn
by construction in the non-relativistic limit. This is so because we have implicitly used this equation (via the spatial part of the
equations of motion) to define
what we mean with non-relativistic limit. In fact one can read this backwards to see that fulfilling
the Euler equation means $\chi - V$ is a constant in this limit.

\subsection{AdS black holes}

But empty AdS space does not get us very far.
The most commonly used metric for the applications of the gauge-gravity duality is the charged AdS planar black hole. Let us first
leave aside the charge. In $d$+1 dimensions, this metric has
the following line element:
\beqn
{\rm d}s^2 = - \frac{L^2}{z^2} \left( 1 - \frac{z^d}{z_0^d}\right) {\rm d}t^2 + \frac{L^2}{z^2} \left( 1 - \frac{z^d}{z_0^d}\right)^{-1} {\rm d}z^2 + \frac{L^2}{z^2}  \sum_{i=1}^{d-1} {\rm d} x^i {\rm d} x^i  ~. \label{planar}
\eeqn
The horizon is located at $z = z_0$. The
temperature of the black hole can be computed from the surface gravity of the horizon and is proportional to $1/z_0$. 

Note that
the sum over the transverse coordinates is Cartesian, which is why the black hole is said to be planar. The horizon
is translationally invariant in the $x^i$-directions. The spherically symmetric solution looks the same as the planar one except that the line
element on the spatial slices is that of a sphere (in the appropriate number of dimensions). One uses
the planar solution so that the boundary of this AdS space is a flat space-time which accurately describes
the setting that one finds in the laboratory. The planar black hole solutions are peculiar and mathematically
very interesting because there is no similar solution with non-compact event horizon
in asymptotic Minkowski-space. The cosmological constant term in AdS space is necessary for these
solutions to exist. 

\subsubsection{Method 1: Shifting}

To move on, the metric (\ref{planar}) first has to be brought into a form suitable to find an analog system.
If we want the spatial subspaces to be conformally flat, we can do the same transformation
as previously to generate a non-vanishing shift vector in the $g_{ti}$-coordinates. Again we
make the ansatz $t=t'+f(z)$, where now
\beqn
f(z) = \int {\rm d}z \left( \frac{z}{z_0} \right)^{d/2} \left( 1 - \frac{z^d}{z^d_0} \right)^{-1} \quad,
\eeqn
leads to the metric
\beqn
{\rm d}s^2 = - \frac{L^2}{z^2} \left( 1 - \frac{z^d}{z_0^d}\right) {\rm d}t'^2 - \frac{L^2}{z^2} \left( \frac{z}{z_0}\right)^{d/2} {\rm d}t' {\rm d}z + \frac{L^2}{z^2} {\rm d}z^2 + \frac{L^2}{z^2} \sum_{i=1}^{d-1} {\rm d} x^i {\rm d} x^i ~,\label{planarod}
\eeqn
which then has a non-zero shift vector.
Next we rescale as previously the time-like coordinate by a constant factor $\kappa$ for later convenience:
 \beqn
{\rm d}s^2 = - \frac{L^2}{z^2} \left( 1 - \frac{z^d}{z_0^d}\right) \kappa^2 {\rm d}t'^2 - \frac{L^2}{z^2} \left( \frac{z}{z_0}\right)^{d/2} \kappa {\rm d}t' {\rm d}z + + \frac{L^2}{z^2} {\rm d}z^2 + \frac{L^2}{z^2} \sum_{i=1}^{d-1} {\rm d} x^i {\rm d} x^i ~. \label{planarodk}
\eeqn

From the spatial diagonal elements we can then read off by comparing with (\ref{gdown}) that $\rho_0/(c m^2 a^2) = L^2/z^2$, and from the off-diagonal elements $(v_0)_z^2=v_0^2 = \kappa (z/z_0)^{(d/2)}$. 
From the $g_{tt}$-component we then get $c^2=\kappa^2$, and so 
\beqn
\rho_0 v_0 \propto z^{d/2-2}~.
\eeqn
Since we have a planar symmetry this means that the gradient is just $\partial_z$ and the Euler-equation is fulfilled if and only if $d=4$. And so we find that, unlike the case for Schwarzschild black holes in $3+1$ dimensions, the $4+1$ dimensional planar black hole can be
identically expressed by an analog gravity system, without additional conformal prefactors. For this both the dimension
of the black hole is relevant as well as its planar symmetry, which is only possible in asymptotic AdS-space. 

We can calculate the force-density for this system using the Euler equation
\beqn
F_z = - \rho_0 (v_0)_z \partial_z (v_0)_z - c^2 \partial_z \rho_0 =F_{\rm AdS} - 2 \kappa^3 (maL)^2 \frac{z}{z_0^4} = \gamma(z) F_{\rm AdS} ~. \label{force1}
\eeqn
Note that these relations do not fix a particular value of the speed of sound. They merely fix the relation between
the speed of sound and the fluid velocity, the two of which have to be equal at the horizon. As noted above, this
force is identically canceled by the respective potential term from the Lagrangian.

\subsubsection{Method 2: Rescaling}

The planar
symmetry allows one to find an even easier analog system by starting again with the coordinate system (\ref{planar}) and 
just rescaling the asymptotic coordinate in (\ref{planar}) to $\tilde{z} = \tilde{z}(z)$. All we have to do is demand that
the new metric is conformally flat in the spatial subspaces. This means 
\beqn
\tilde z = \int_0^z \frac{{\rm d} z'}{\sqrt{|\gamma(z')}|} ~ \label{tildez},
\eeqn
which gives the metric
\beqn
{\rm d}s^2 = - \frac{L^2}{z(\tilde z)^2} \left( 1 - \frac{z(\tilde z)^d}{z_0^d}\right) \kappa^2 {\rm d}t^2 + \frac{L^2}{z(\tilde z)^2}  {\rm d} \tilde z^2 + \frac{L^2}{z(\tilde z)^2}  \sum_{i=1}^{d-1} {\rm d} x^i {\rm d} x^i  ~. \label{planarres}
\eeqn
We here have once again rescaled the time-coordinate with a constant $\kappa$.

Then we have a non-relativistic analog metric with 
\beqn
(v_0)_i = 0~,~ \rho_0 = \kappa (Lma)^2 \frac{ \sqrt{\gamma(\tilde z)}}{z(\tilde z)^2} ~,~c = \kappa \sqrt{\gamma(\tilde z)}~.  \label{nranalog}
\eeqn
The continuity equation is trivially fulfilled in this case because the metric has no time-dependence and $\rho_0 v_0^i = 0$. This rescaling of the asymptotic coordinate 
is only possible for planar symmetry because otherwise the line-element of the subspaces would no longer be conformally flat. 

Again we
can calculate the necessary force-density to obtain the correct speed of sound:
\beqn
F_z = 2 \kappa^3 (Lma)^2 \frac{\sqrt{\gamma}}{z^3} \left(\gamma+ \frac{d}{4} \left( \frac{z}{z_0}\right)^{d} \right)~.
\eeqn
We have to keep in mind though that we previously transformed $z$ into $\tilde z$, so in the lab coordinates we actually have
\beqn
F_{\tilde z} = \sqrt{\gamma} F_z =  2 \kappa^3 (Lma)^2 \frac{\gamma}{z^3} \left(\gamma+ \frac{d}{4} \left( \frac{z}{z_0}\right)^{d} \right)~. \label{Fgamma}
\eeqn
For $d=4$ the factor in the large brackets is just equal to one and the force is the same as in the other coordinate system (\ref{force1}). The difference
between the both cases is the density and velocity profile, and the speed of sound. In the former case, the speed of sound is
constant while the speed of the background changes, so that a sonic horizon occurs when the perturbations on the fluid
cannot travel fast enough to escape. In the rescaled case, it is instead the speed of sound that changes. The method
of rescaling is added here for completeness because it succeeds in bringing the metric of the AdS planar black hole
into the desired form, but we will not use it in the following.

\subsection{Charged AdS planar black holes}

The metric of the charged planar black hole in asymptotic AdS can, like the metric of the uncharged black hole, be written in the form
\beqn
{\rm d}s^2 = - \frac{L^2}{z^2} \tilde \gamma(z) {\rm d}t^2 + 
\frac{L^2}{z^2} \tilde \gamma(z)^{-1} {\rm d}z^2 + \frac{L^2}{z^2}  \sum_{i=1}^{d-1} {\rm d} x^i {\rm d} x^i  ~. \label{planarcharge}
\eeqn
where now \cite{Hartnoll:2009sz}
\beqn
\tilde \gamma(z) =  1 - (1+\alpha^2) \left( \frac{z}{z_0}\right)^d + \alpha^2 \left(\frac{z}{z_1}\right)^{2(d-1)}~ \label{gammacharge}
\eeqn
with
\beqn
\alpha^2 = z_0^2 \mu^2 \frac{d-2}{d-1} ~,
\eeqn
and $\mu$ the chemical potential on the boundary. The electromagnetic potential in the AdS background
that gives rise to this metric is \cite{Hartnoll:2009sz}
\beqn
A_t = \mu \left( 1 - \left( \frac{z}{z_0} \right) ^{d-2} \right)~. \label{A}
\eeqn

The previously made observation that we can rescale the $z$-coordinate to make the subspaces conformally flat immediately
allows us to find an analog metric for the charged case. With $\tilde{z}$ defined by Eq (\ref{tildez}) where $\gamma(z)$ is now
replaced with $\tilde \gamma(z)$ given by Eq (\ref{gammacharge}), we still have a non-relativistic analog metric by the identifications (\ref{nranalog}). 
We can then go on and calculate the force-density again using Eq (\ref{force1}), but one then sees that in this case 
there is no obvious relation between the force in the charged and uncharged case.
This seems unintuitive. One would expect that turning on an external field for a fluid (or material) with suitable electromagnetic
response should produce the family of solutions without adding another force. This leads us to look again at the
metric with the off-diagonal terms and the non-vanishing shift or velocity field respectively. Even though it is more complicated, the
necessity of rescaling the asymptotic coordinate does not make much physical sense.

Now recall that we want to reproduce the wave-equation of the charged scalar field in the AdS charged planar
black hole background. The potential that the excitation has to couple to is given by Eq (\ref{A}), and thus has to appear in the wave-equation. 
The question we have to address now is whether adding this potential also gives rise to the correct metric. 

For this we first couple
the background field to the gauge field by replacing $\chi \to \tilde \chi$, where now
\beqn
\tilde \chi = \eta^{\mu \nu} D_\mu \theta D_\nu \theta^*~.
\eeqn
Then we add the term for the free gauge field \cite{Novello:1999pg} so that
\beqn 
{\cal L}= {\cal L}(\tilde \chi (D \theta, D\theta^*) - V(\theta, \theta^*, \vec x, t) - \frac{1}{4} F^2~)~.
\eeqn 
Here  $F$ is the field-strength tensor as usual\footnote{There is no risk of confusion with the force because the field-strength tensor will
either have two indices or its trace have none, whereas the force has one index.}, and as previously $D = \partial + {\rm i} e A$ is the gauge-covariant derivative.
Note that while this is the simplest way to create  a 
gauge-invariant Lagrangian, it is not the only way because we could have different functional forms for the matter and the
gauge sector. Different forms of the Lagrangian correspond to different hydrodynamic and electromagnetic properties of the medium.
We will see however that this simple choice suits our needs. The reader be warned that since we are dealing with
excitations rather than elementary particles, the charge of the excitations may not be the elementary charge. 

The Maxwell equations of the fluid coupled to the field then read
\beqn
e \frac{\partial {\cal L}}{\partial \tilde \chi}\tilde J^\mu = \partial_\nu \left( \frac{\partial {\cal L}}{\partial \tilde \chi}  F^{\nu \mu} \right)~ \label{Maxwell}
\eeqn
with
\beqn
\tilde J_\nu = {\rm i} ( \theta_0 D_\nu \theta_0^* - \theta_0^* D_\nu \theta_0 )~. \label{current}
\eeqn

We have neglected here that a charged excitation will also create a perturbation of the electromagnetic field, which is why
only the $\theta_0$ appears in (\ref{current}). Then the 
derivation of the effective metric in Eqs (\ref{beginderiv}) to (\ref{endderiv}) remains the same, except that we have
to add all the complex conjugate terms because the scalar field is now complex, and we have to replace all partial
derivatives with covariant derivatives. This means in particular that the velocities and densities that appear in the
metric are not the physical ones because to reproduce the nice form of the metric we have to identify
\beqn
T_{\mu \nu} = (\tilde p_0+\tilde \rho_0)\tilde u_\mu \tilde u^*_\nu +p_0 \eta_{\mu \nu} - \frac{\partial \cal L}{\partial \tilde \chi} F_\mu^{\;\alpha} F_{\alpha \nu} ~, \label{Tmunu}
\eeqn
where
\beqn
\tilde u_\mu = \frac{D_\nu \theta}{\sqrt{\chi - e \tilde J_\alpha A^\alpha}}~,~ p_0 = {\cal L}~,~ \tilde \rho_0 = 2 (\chi - e \tilde J_\alpha A^\alpha)  \frac{\partial {\cal L}}{\partial \tilde \chi} - \tilde p_0 ~. \label{ident3}
\eeqn

It has been demonstrated in \cite{Novello:1999pg} that if one only considers the electromagnetic field and a
perturbation over it, then this perturbation will feel an effective metric from the electromagnetic background field,
which can be derived from (\ref{ident2}) the same way we previously derived the effective metric generated by
the background fluid. However, we are here interested in perturbations of the fluid, not of the electromagnetic field.

With the quantities (\ref{ident3}), the metric then again takes the form
\beqn
g_{\mu \nu} = \tilde c^{-1}  \left(  \frac{\tilde \rho_0+ \tilde p_0}{\chi - e \tilde J_\alpha A^\alpha} \right)  \left( \eta_{\mu\nu} + \left(1- \tilde c^2 \right) \tilde u_\mu \tilde u_\nu \right) ~, \label{endderiv2}
\eeqn
where now 
\beqn
\frac{1}{\tilde c^2} :=  \frac{2 \tilde \chi \partial^2 {\cal L}/\partial \tilde \chi^2 + \partial {\cal L}/\partial \tilde \chi}{\partial{\cal L}/\partial \tilde \chi} ~.
\eeqn

Having found the effective metric of the fluid coupled to the electromagnetic field, we continue to again 
bring the metric (\ref{planarcharge}) in the off-diagonal form
with the by now familiar transformation $t = t' + f(z)$ with $f(z)$ as in Eq (\ref{genf}) but with $\tilde \gamma$ from Eq (\ref{gammacharge}). The metric of
the charged planar black hole then reads
\beqn
{\rm d}s^2 = - \frac{L^2}{z^2} \tilde \gamma(z) \kappa^2 {\rm d}t'^2 - \frac{L^2}{z^2} \sqrt{\tilde \gamma(z) -1} \kappa {\rm d}t' {\rm d}z + \frac{L^2}{z^2} {\rm d}z^2 + \frac{L^2}{z^2} \sum_{i=1}^{d-1} {\rm d} x^i {\rm d} x^i ~, \label{planarodc}
\eeqn
where we have once again rescaled the time-coordinate by a constant factor $\kappa$. From this we can read off right away that $\partial {\cal L}/\partial \tilde \chi = \tilde c L^2/z^2$. The transformation of coordinates also transforms the vector potential $A_\kappa$ and generates a non-zero $A_z$. However, since this 
term is a function of $z$ only it is a pure gauge term and not of physical relevance, so we will disregard it. With that one then obtains the current from the Maxwell equations (\ref{Maxwell}) as
\beqn
\tilde J_t = \frac{z^2}{\tilde c} \partial_z \left( \frac{\tilde c}{z^2} \partial_z A_t \right) \label{J}~.
\eeqn

In the uncharged case it was the continuity equation that told us which metric can be realized in a gravitational analog,
and so we turn again to the continuity equation which, in the time-independent, planar and non-relativistic limit just
takes the form $\partial_z (\rho_0 v_0^z ) = 0$. Alas, the quantities appearing in this equation are the physical density
and velocity and not the gauge-covariant ones that we have used for convenience. We can convert one into the
other by comparing Eq (\ref{ident}) with Eq (\ref{ident3}). Since $A_i =0$ we find
\beqn
\tilde \rho_0 = \rho_0 \sigma^2  ~,~ \tilde p_0 = p_0 \sigma^2~,~\tilde v_0^z = \frac{v_0^z}{\sigma} \quad \mbox{with} \quad \sigma^2 := \tilde \chi/\chi~.
\eeqn

We then also take the non-relativistic limit of the metric because the analog metric which we had identified in the uncharged
case was also the one from the non-relativistic limit. For this we will have to assume that also the coupling term to the electromagnetic
field is much smaller than the trapping potential. If the field was too strong, it would increase the velocity of the background fluid
so that we could no longer use the non-relativistic limit, and the approximation would not make sense. 
We have then as before $\chi \to (m a)^2$ (note no tilde) and
\beqn
g_{\mu \nu} (t, {\vec x}) &\to& \left( \frac{\rho_0}{\tilde c (ma)^2} \right)^{\frac{2}{n-1}}
\left( \begin{array}{cc}
- \kappa^2 (\tilde c^2- v_0^2/\sigma^2) & - \kappa v_0^j/\sigma \\
- \kappa v_0^i/\sigma & \delta_{ij}   \end{array} \right) \label{gdowndcharged} ~,
\eeqn
where
\beqn
\sigma^2 \to 1 -  e \tilde J_t A_t/(ma)^2 ~.
\eeqn
Then we can then as previously read off
\beqn
\tilde c &=& \kappa~,~\rho_0 = \kappa (maL)^2 \frac{1}{z^2}~, \\
 (v_0)_z &=& \sigma \kappa   \sqrt{ (1+\alpha^2) \left( \frac{z}{z_0}\right)^d - \alpha^2 \left(\frac{z}{z_0}\right)^{2(d-1)}}~.
\eeqn
Knowing that $\tilde c$ is a constant, we can now calculate the current from Eq. (\ref{J}) to
\beqn
\tilde J_t = \mu z_0^{2-d} (d-2)(d-5) z^{d-4}~,
\eeqn
with some exceptions in case a derivative vanishes, but these cases don't matter for us as we will  see shortly. 

The continuity equation then has, again in the non-relativistic limit, indeed the usual form
\beqn
\partial_z \left( \rho_0 v_0^z \right) = 0 ~.
\eeqn
There is no additional source term from the electromagnetic field because the field is static.
One now convinces oneself easily that the continuity equation is fulfilled if and only if $d=4$ and 
\beqn
z_0^4 (ma)^2= e^2/3 ~. 
\eeqn
In this case $\tilde J_t$ is constant and $\partial {\cal L}/\partial A \propto \partial {\cal L}/\partial \chi$. 
Since $\tilde J_i =0$, this means we have a non-conducting medium whose charge density is
proportional to the matter density. 

The surprising finding here is not that the continuity equation can only be fulfilled in $d=4$. We already knew this because the uncharged case only
works in $d=4$, so any other number of dimensions would not work. The surprising finding is that the gravitational analogy still works with the electric field added. Just turning on an electric field
can of course not lead to violations of the continuity equation, but it was not a priori obvious that the generated effective metric would also
be 
the metric that would be obtained from gravitational response to the electric field via Einstein's field equations.

\section{Mass and Potential}
\label{mass}

So far we have shown that it is possible to find a gravitational analog for the propagation of a charged scalar perturbation in
the background metric of a charged, planar, AdS black hole. We have not yet looked at the mass of these fields. As previously
noted, we will need a potential with a minimum around which the background field can condense. We will take that
to be the simplest potential that we can come up with for this purpose which is the vanilla Mexican hat
\beqn
V(\theta,\theta^*) =  m^2 \theta \theta^* - \frac{\lambda^2}{2} (\theta \theta^*)^2 ~.\label{potgauge}
\eeqn
For this potential then 
\beqn
\langle \theta_0 \rangle = m/\lambda~,~V(\langle \theta_0 \rangle) = \frac{1}{2} \frac{m^4}{\lambda^2}~,
\eeqn
and so $2 a^2 = m^2/\lambda$, to connect to the earlier used parameters (see Eq. (\ref{nrl}). 

However, since the potential (\ref{potgauge}) is gauge-invariant, the mass-term that is generated for the perturbation vanishes. If we do not want this
mass term to vanish, the easiest way to get one is to modify the potential so that it contains a small term which does not respect
the gauge invariance (but is still hermitian), for example
\beqn
V(\theta,\theta^*,\vec x, t) =  (m^2 - \nu (\vec x, t)^2)  \theta \theta^* - \frac{\lambda^2}{2} (\theta \theta^*)^2 + 
\frac{\nu (\vec x, t)^2}{2} \left( \theta \theta + \theta^* \theta^* \right) \label{potgauge2}. 
\eeqn
This potential still has the same minimum for $\theta_0$ as (\ref{potgauge}), but the terms with $\nu$ will now generate a mass-term for the excitation.
Since this mass-term for the excitation does, in the non-relativistic limit, not affect the equations of motions of the background field, adding
a mass to the excitation will not lead to additional constraints on the fluid, provided that the gauge-invariance violating contribution is
sufficiently small. 

Let us then have a closer look at the effective\footnote{Note that in our notation, the effective mass does not contain the term proportional to $A_t^2$ that is generated by the electric field. This term is contained in the covariant derivative.} mass of the perturbation, Eq. (\ref{meff}). This term is the derivative with respect to $\theta$ at the
minimum of the action where the background field fulfills the equations of motion. The first term in this expression, for
the case of two derivatives with respect to $\theta$ can
be rewritten as
\beqn
\frac{\partial {\cal L}^2}{\partial \theta^2}\Bigg|_{\theta = \theta_0} =  \left( \frac{\partial^2 V}{\partial \theta^2} \frac{\partial {\cal L}}{\partial \chi} + \left( \frac{\partial V}{\partial \theta} \right)^2 \frac{\partial^2 {\cal L}}{\partial \chi^2} \right) \Bigg|_{\theta = \theta_0}~,
\eeqn
where we have emphasized that we first have to take the derivative and then enter the value of the background field. The second term on the right
side of this expression vanishes then by assumption in the limit that we work in at $\theta= \theta_0$. The same conclusion applies for all combinations of $\theta$ and $\theta^*$. 

The second term in the effective mass, after some rewriting and summing over all combinations of $\theta$ and $\theta^*$ (taking into account that each $\theta^*$ will bring in a minus for the perturbation), takes the form
\beqn
\partial_\nu  \left( \left[ \frac{\partial V}{\partial \theta} - \frac{\partial V}{\partial \theta^*}\right] 
\frac{\partial^2 {\cal L}}{\partial \chi^2} \left( \partial^\nu \theta^* - \partial^\nu \theta \right) \right) \Bigg|_{\theta = \theta_0}
\eeqn
So this term doesn't contribute to the effective mass either, and we only have the first term that is proportional to the second derivatives of the potential $V$. (Which, in hindsight, justifies the expectation that Goldstone's theorem isn't affected by the form of the Lagrangian.)

We can
then work out the effective mass term  from (\ref{potgauge}) as 
\beqn
m_{\rm eff}^2 = \frac{1}{\sqrt{-g}} 2 \nu(z)^2  \frac{\partial L}{\partial \chi} ~.
\eeqn
If we want the effective mass to be position-independent, then $\nu(z)^2 \sim 1/z^2$. If we on the other hand chose $\nu$ to be a 
constant then the effective mass will be proportional to $z^2$. Since $\nu$ by construction only appears in the wave-equation for
the perturbation, it is not affected by the previous considerations that concern the equations of the background field. Just why this potential scales this way is a question that
could only be addressed by a more complete model that takes into account the origin of $\nu$ and the origin of the mass of
the scalar field in AdS space.

\section{Discussion}

In the previous sections, we have focused on the case $n=3, d=4$. One sees however from the treatment of the non-charged black hole that the continuity equation
will be fulfilled whenever the powers of $z$ in the density cancel the powers of $z$ in the velocity. Since $\rho_0^{2/(n-1)} \propto 1/z^2$ and 
$v_0^z \propto z^{d/2}$ this will be the case whenever $2(n-1) = d$. This leaves open the question though whether the extension to the charged black hole
also works in all these cases.

One further expects that if the analogy is a general property of the systems under investigation, then there should
be a relativistic completion of the non-relativistic case that we have considered here. If one makes a general ansatz, one finds that
the method of shifting alone does not lead to a non-relativistic analog metric, and one thus has to work with a combination of shifting and rescaling.
Furthermore, we have dealt here only with the classical analog metric. It would be interesting to see whether the quantum case also works.

The treatment discussed here does not take into account backreaction in any sort. In principle this should be possible to do using the
perturbative ansatz and taking along higher powers. It remains to be seen then whether the backreaction from the perturbations of
the fluid on the background act like the backreaction of matter on the space-time geometry. This is not a priori impossible since we here have
 a system with many symmetries and do not have to deal with the Einstein equations in full generality.

Since we are here working with a projection onto a slice of AdS space, the analog condensed matter system cannot reproduce all the degrees of
freedom of the bulk theory and does not allow the propagation of all modes. It will thus also not in general reproduce the complete
boundary theory, but only a subset with suitable symmetry. It would then be interesting to see if a connection can be made between the
Lagrangian approach considered here and the string-theoretical origin of the gauge-gravity correspondence. It has for example
been noted previously that Lagrangians of the general type used here describe the dynamics of certain p-branes \cite{Gibbons:2000xe,Bilic:2006cp}.

It is curious to note that the reason the charged case works is that in asymptotic $4+1$ dimensional AdS space, $\tilde \gamma -1 \propto z^4 A_t$. This
however is only so because $A_t$ is gauged so to have a constant term, which again is chosen to reproduce the chemical potential on the
boundary. 

It has been previously suggested to experimentally observe the AdS/{\sc CMT} correspondence in lower-dimensional systems using
graphene \cite{holo,holo2,Chen:2012uc}, but
the general approach presented here suggests that there might exist a deeper relationship between certain condensed matter systems. 

Finally, there arises of course the question whether it is possible to realize the analog system for planar black holes experimentally, in which
case one has to take into account further approximations, notably finite size corrections. For an experimental realization one will have to find a system that 
reproduces the required density and velocity profile and, in the charged case, the suitable electromagnetic response. Since the acoustric
metric considered here is only one particular type of analog metrics,  one can also study other systems, for example dielectric matter \cite{Gordon,optic1} that
could provide very different possibilities to realize the planar black hole metric. It might prove challenging to find a suitable analog gravity 
system. But the here demonstrated existence of a Lagrangian for a fluid that gives rise to the desired metric has the benefit that it allows one to search for a formulation of the duality
directly on the Lagrangian level. This, when found, would implicitly provide a formulation of the AdS/CFT duality on the Lagrangian level,
which is so far missing.

Let us now step back for a moment and reflect on what we have found. 

We have found 
that the metrics which appear as gravitational duals to condensed matter systems on the AdS 
boundary are just of the right form to also fulfill the fluid equations of motion of an analog
metric.  There was a priori no reason for this to be the 
case. How come that the same gravitational system allows for two different descriptions
in form of condensed matter systems? Unlike the case for the Schwarzschild black hole, the scaling for 
these metrics fits without the need to add an additional conformal factor. The identification of the
system on the $3+1$ dimensional boundary only works if the AdS subspace is also $3+1$
dimensional. The scaling still matches when we turn on an electric field. This may be
a curious coincidence or it may be a consequence of a deeper underlying relation the
full extend of which is yet to be discovered.

This matching between the systems then allows one to identify physical observables
of the systems. To look into this one would have to specify the Lagrangian in more detail than
we have done here, but a simple example can be read off from the metric without much
effort. That is that the temperature of the system on the AdS boundary is proportional
to the speed of sound of the analog system because both can be identified with the the position
of the horizon. The closer the horizon is to the center of AdS space, the higher the temperature
on the boundary. The larger the speed of sound in the analog system, the better perturbations
can escape the sonic horizon, thus moving it closer to the center. 

However, we have dealt here only with a fixed background, which is a necessary but
not a sufficient condition to allow us to truly identify the two systems with each other. 
To complete the identification we have begun here, the backreaction of the perturbations
on the fluid background in the analog system has to be studied, and it has to be seen
whether they reproduce the fully coupled system between the metric components and
the fields propagating in it. There here found relation is interesting though on its own
right by demonstrating that these asymptotic AdS spaces exist as analog
systems and that in the probe approximation also the propagation of perturbations
can be matched.

\section{Brief Summary}

We have demonstrated that charged planar black holes in asymptotic $4+1$ dimensional AdS space have a gravitational
analog system which is a charged, nonconducting fluid. The density of the fluid scales like the AdS conformal factor $1/z^2$,
and the charge density is proportional to the mass density. 

The analog metric we have studied here is not that of the full $4+1$ dimensional AdS space, but that of a $3+1$ dimensional
projection, where the projection is onto one of the spatial coordinates that are perpendicular to the AdS asymptotic coordinate.
This projection is relevant for the identification to work, not to mention that would be hard to realize a $4+1$ dimensional
fluid in the laboratory. 

It must be emphasized that in contrast to the well-studied case
of the Schwarzschild black hole, the analog system of the (projection of the) planar black hole in $4+1$ dimensions 
does not merely produce a metric that is conformal to that of the black hole, but in the AdS-case it is actually 
the same metric with the correct scaling already. The reason for this is the planar symmetry of the system,
which is only possible in asymptotic AdS space.

In the end the central question is whether the here discussed relation between two entirely different condensed matter
systems is a mere coincidence, or whether it not hints at a deeper underlying principle.

\section*{Acknowledgements} 

I am grateful for helpful feedback from Neven Bili\'c, Mirah Gary, Valentina Giangreco Puletti, Stefano Liberati, Stefan Scherer, L\'arus Thorlacius and 
Silke Weinfurtner.


\begin{thebibliography}{99}
\small{ 

\bibitem{Unruh:1980cg} 
  W.~G.~Unruh,
  Phys.\ Rev.\ Lett.\  {\bf 46}, 1351 (1981).

\bibitem{Barcelo:2005fc} 
  C.~Barcelo, S.~Liberati and M.~Visser,
  Living Rev.\ Rel.\  {\bf 8}, 12 (2005)
  [Living Rev.\ Rel.\  {\bf 14}, 3 (2011)]
  [gr-qc/0505065].

\bibitem{Visser:1997ux} 
  M.~Visser,
  Class.\ Quant.\ Grav.\  {\bf 15}, 1767 (1998)
  [gr-qc/9712010].
\bibitem{Garay:1999sk} 
  L.~J.~Garay, J.~R.~Anglin, J.~I.~Cirac and P.~Zoller,
  Phys.\ Rev.\ Lett.\  {\bf 85}, 4643 (2000)
  [gr-qc/0002015].

\bibitem{Barcelo:2000tg} 
  C.~Barcelo, S.~Liberati and M.~Visser,
  Class.\ Quant.\ Grav.\  {\bf 18}, 1137 (2001)
  [gr-qc/0011026].

\bibitem{Volovik:2000ua} 
  G.~E.~Volovik,
  Phys.\ Rept.\  {\bf 351}, 195 (2001)
  [gr-qc/0005091].

\bibitem{Weinfurtner:2004mu} 
  S.~E.~C.~Weinfurtner,
  Gen.\ Rel.\ Grav.\  {\bf 37}, 1549 (2005)
  [gr-qc/0404063].

\bibitem{Jain:2007gg} 
  P.~Jain, S.~Weinfurtner, M.~Visser and C.~W.~Gardiner,
  Phys.\ Rev.\ A {\bf 76}, 033616 (2007)
  [arXiv:0705.2077 [cond-mat.other]].

\bibitem{Lin:2012ft} 
  C.~Y.~Lin, D.~S.~Lee and R.~J.~Rivers,
  J.\  Phys.\  : Condens.\  Matter {\bf 25}, 404211 (2013)
  [arXiv:1205.0133 [gr-qc]].

\bibitem{Bilic:2013qpa} 
  N.~Bilic and D.~Tolic,
  Phys.\ Rev.\ D {\bf 88}, 105002 (2013)
  [arXiv:1309.2833 [gr-qc]].



\bibitem{Weinfurtner:2010nu} 
  S.~Weinfurtner, E.~W.~Tedford, M.~C.~J.~Penrice, W.~G.~Unruh and G.~A.~Lawrence,
  Phys.\ Rev.\ Lett.\  {\bf 106}, 021302 (2011)
  [arXiv:1008.1911 [gr-qc]].

\bibitem{ex1} J.~Steinhauer, Nature Physics {\bf 10}, 864–869 (2014) [arXiv:1409.6550 [cond-mat.quant-gas]].

\bibitem{Maldacena:1997re} 
  J.~M.~Maldacena,
  Adv.\ Theor.\ Math.\ Phys.\  {\bf 2}, 231 (1998)
  [hep-th/9711200].
\bibitem{Witten:1998qj} 
  E.~Witten,
  Adv.\ Theor.\ Math.\ Phys.\  {\bf 2}, 253 (1998)
  [hep-th/9802150].
\bibitem{Gubser:2002tv} 
  S.~S.~Gubser, I.~R.~Klebanov and A.~M.~Polyakov,
  Nucl.\ Phys.\ B {\bf 636}, 99 (2002)
  [hep-th/0204051].


\bibitem{Hartnoll:2008kx} 
  S.~A.~Hartnoll, C.~P.~Herzog and G.~T.~Horowitz,
  JHEP {\bf 0812}, 015 (2008)
  [arXiv:0810.1563 [hep-th]].

\bibitem{Hartnoll:2009sz} 
  S.~A.~Hartnoll,
  Class.\ Quant.\ Grav.\  {\bf 26}, 224002 (2009)
  [arXiv:0903.3246 [hep-th]].

\bibitem{Horowitz:2010gk} 
  G.~T.~Horowitz,
  Lect.\ Notes Phys.\  {\bf 828}, 313 (2011)
  [arXiv:1002.1722 [hep-th]].

\bibitem{Musso:2014efa} 
  D.~Musso,
  arXiv:1401.1504 [hep-th].
\bibitem{Polchinski:2014mva} 
  J.~Polchinski,
  arXiv:1412.5704 [hep-th].

\bibitem{Horowitz:2012ky} 
  G.~T.~Horowitz, J.~E.~Santos and D.~Tong,
  JHEP {\bf 1207}, 168 (2012)
  [arXiv:1204.0519 [hep-th]].

\bibitem{Horowitz:2013jaa} 
  G.~T.~Horowitz and J.~E.~Santos,
  arXiv:1302.6586 [hep-th].

\bibitem{ElShowk:2011ag} 
  S.~El-Showk and K.~Papadodimas,
  JHEP {\bf 1210}, 106 (2012)
  [arXiv:1101.4163 [hep-th]].

\bibitem{Fitzpatrick:2012cg} 
  A.~L.~Fitzpatrick and J.~Kaplan,
  JHEP {\bf 1302}, 054 (2013)
  [arXiv:1208.0337 [hep-th]].


\bibitem{Bilic:2014dda} 
  N.~Bili\'c, S.~Domazet and D.~Tolić,
  arXiv:1410.0263 [hep-th].

\bibitem{Bilic:1999sq} 
  N.~Bilic,
  Class.\ Quant.\ Grav.\  {\bf 16}, 3953 (1999)
  [gr-qc/9908002].


\bibitem{Visser:2010xv} 
  M.~Visser and C.~Molina-Paris,
  New J.\ Phys.\  {\bf 12}, 095014 (2010)
  [arXiv:1001.1310 [gr-qc]].


\bibitem{PG1} P.~ Painlev\'e, 
C.\ R.\ Acad. Sci. (Paris) 173, 677–680 (1921).
\bibitem{PG2} A.~Gullstrand, 
Arkiv.\ Mat.\ Astron.\ Fys.\ 16(8), 1–15 (1922).
\bibitem{PG3} G.~ Lemaitre, Annales de la Soci\'et\'e Scientifique de Bruxelles A53: 51–85 (1933).

\bibitem{Novello:1999pg} 
  M.~Novello, V.~A.~De Lorenci, J.~M.~Salim and R.~Klippert,
  Phys.\ Rev.\ D {\bf 61}, 045001 (2000)
  [gr-qc/9911085].


\bibitem{Gibbons:2000xe} 
  G.~W.~Gibbons and C.~A.~R.~Herdeiro,
  Phys.\ Rev.\ D {\bf 63}, 064006 (2001)
  [hep-th/0008052].

\bibitem{Bilic:2006cp} 
  N.~Bilic, G.~B.~Tupper and R.~D.~Viollier,
  J.\ Phys.\ A {\bf 40}, 6877 (2007)
  [gr-qc/0610104].

\bibitem{Gordon} W.~Gordon, Ann.\ Phys.\ (Leipzig) {\bf 72}, 421 (1923).

\bibitem{optic1} U.~Leonhardt,
Phys.\ Rev.\ A {\bf 62}, 012111 (2000) [physics/0001064].

\bibitem{holo} D.~ V.~ Khveshchenko, EPL 104, 47002 (2013) [arXiv:1305.6651 [cond-mat.str-el]].

\bibitem{holo2} G.~W.~Semenoff, AIP Conf.\ Proc.\ 1483, 305 (2012).
\bibitem{Chen:2012uc} 
  P.~Chen and H.~Rosu,
  Mod.\ Phys.\ Lett.\ A {\bf 27}, 1250218 (2012)
  [arXiv:1205.4039 [gr-qc]].

}

\end{thebibliography}
\end{document}